\documentclass[twocolumn,aps,prl,superscriptaddress,showpacs,tightenlines]{revtex4}
\usepackage{amsmath}
\usepackage{amsfonts}
\usepackage{graphicx}
\usepackage{epsfig}
\usepackage{color}
\usepackage[colorlinks,citecolor=blue]{hyperref}

\begin{document}

\title{$\mathcal{PT}$-Symmetry-Breaking Chaos in Optomechanics}

\author{Xin-You L\"{u}}
\email{xinyoulu@gmail.com}
\affiliation{Wuhan National Laboratory for Optoelectronics and School of physics, Huazhong University of Science and Technology, Wuhan 430074, PeopleÕs Republic of China}
\author{Hui Jing}
\affiliation{Department of Physics, Henan Normal University, Xinxiang 453007, PeopleÕs Republic of China}
\author{Jin-Yong Ma}
\affiliation{Wuhan National Laboratory for Optoelectronics and School of physics, Huazhong University of Science and Technology, Wuhan 430074, PeopleÕs Republic of China}
\author{Ying Wu}
\email{yingwu2@126.com}
\affiliation{Wuhan National Laboratory for Optoelectronics and School of physics, Huazhong University of Science and Technology, Wuhan 430074, PeopleÕs Republic of China}

\begin{abstract}
We demonstrate a $\mathcal{PT}$-symmetry-breaking chaos in optomechanical system (OMS), which features an ultralow driving threshold. In principle, this chaos will emerge once a driving laser is applied to the cavity mode and lasts for a period of time. 
The driving strength is inversely proportional to the starting time of chaos.
This originally comes from the dynamical enhancement of nonlinearity by field localization in $\mathcal{PT}$-symmetry-breaking phase ($\mathcal{PT}$BP). Moreover, this chaos is switchable by tuning the system parameters so that a $\mathcal{PT}$-symmetry phase transition occurs. This work may fundamentally broaden the regimes of cavity optomechanics and nonlinear optics. It offers the prospect of exploring ultralow-power-laser triggered chaos and its potential applications in secret communication.

\end{abstract}
\maketitle
Cavity optomechanics is a rapidly developing research field exploring the radiation-pressure interaction between the electromagnetic and mechanical systems~\cite{reviews}. Thanks to this nonlinear interaction, the optomechanical system (OMS) provides an alternative platform for implementing many interesting quantum~\cite{Bose1997,Rabl2011,Liao2012,Ludwig2012,Nori2013,Xuereb2012,Nori2015} and classical~\cite{Kippenberg2005,Marquardt2006,Metzger2008,Zaitsev2011,Carmon2007,Meystre2010,Ma2014,Bakemeier2015,Xiong2014} nonlinearity phenomena. In particular, one can strong driving the cavity so that the OMS enters into a regime of self-induced oscillations~\cite{Kippenberg2005,Marquardt2006,Metzger2008,Zaitsev2011},
where the backaction-induced mechanical gain overcomes mechanical loss. Further increasing the strength of the driving laser, chaotic motion emerges both in the optical and mechanical modes requiring no external feedback or modulation~\cite{Carmon2007,Meystre2010,Ma2014,Bakemeier2015}. It is useful for generating random numbers~\cite{Gleeson} and implementing secret information processing~\cite{Sivaprakasam,VanWiggeren,Sciamanna}. However, to apply chaos into the secret communication scheme requiring low-power optical interconnects, the chaos threshold should be reduced dramatically~\cite{Miller,Ellis}. 

The notion of parity-time ($\mathcal{PT}$) symmetry, initially proposed in quantum mechanics~\cite{Bender1998}, attracts recently enormous attention in the field of optics~\cite{Makris2008,Klaiman2008,West2010,Guo2009,Makris2010,Chong2011,Miroshnichenko2011,Lin2011,Regensburger2012,Feng2013,Peng2014,Feng2014,Hodaei2014,Jing2014,Xie2014}. It is known~\cite{Bender1998} that PT symmetric Hamiltonians can exhibit a real eigenvalue spectrum in spite of the fact that they can be non-Hermitian. These systems have the interesting property to undergo an abrupt phase-transition where the system loses its PT-symmetry. At the exceptional point (EP) pairs of eigenvalues collide and become complex. Typically, the transition between PT symmetric phase (real spectrum) to spontaneously broken PT symmetry (complex spectrum) occurs as a parameter in the Hamiltonian (which somehow controls the degree of non-Hermiticity) is varied~\cite{Makris2008,Klaiman2008,West2010}.
This phase transition has been demonstrated in synthetic waveguides and microcavities, and can induce unique optical phenomena including loss-induced transparency~\cite{Guo2009}, power oscillations violating left-right symmetry~\cite{Makris2010}, low-power optical diodes~\cite{Peng2014} and single-mode laser~\cite{Feng2014,Hodaei2014}. A natural question is whether it could influence the chaos dynamics (especially the chaos threshold) significantly. Moreover, the crossover between the $\mathcal{PT}$-symmetry theory and chaos in optomechanics remains largely unexplored, which may substantially advance the fields of cavity optomechanics and nonlinear optics.

Here, we propose a $\mathcal{PT}$-symmetry-breaking chaos~\cite{footnote} by investigating the nonlinear dynamics of a $\mathcal{PT}$-symmetric OMS. In contrast to normal chaos, it is controllable by manipulating $\mathcal{PT}$-symmetry phase transition and allows an ultralow driving threshold, such as the {\it weak-driving} regime, i.e., the driving strength $\Omega_d$ is smaller than the cavity decay rate $\gamma$. Physically, in $\mathcal{PT}$-symmetry-breaking phase ($\mathcal{PT}$BP) the field localization induces the dynamical accumulations of the optical and acoustical energy in the passive cavity, corresponding to an increasing optomechanical nonlinearity with time evolution. This ultimately leads to the result that the chaotic motion can be triggered even if an ultralow-power driving laser lasts for a period of time. As far as we know, this unconventional PT-breaking chaos, featuring an ultralow threshold, is identified for the first time in optomechanics, which may stimulate further investigations and applications in different optical, acoustic, or electric PT-breaking devices, with other types of nonlinearity in the loss element.
\begin{figure}
\includegraphics[width=8cm]{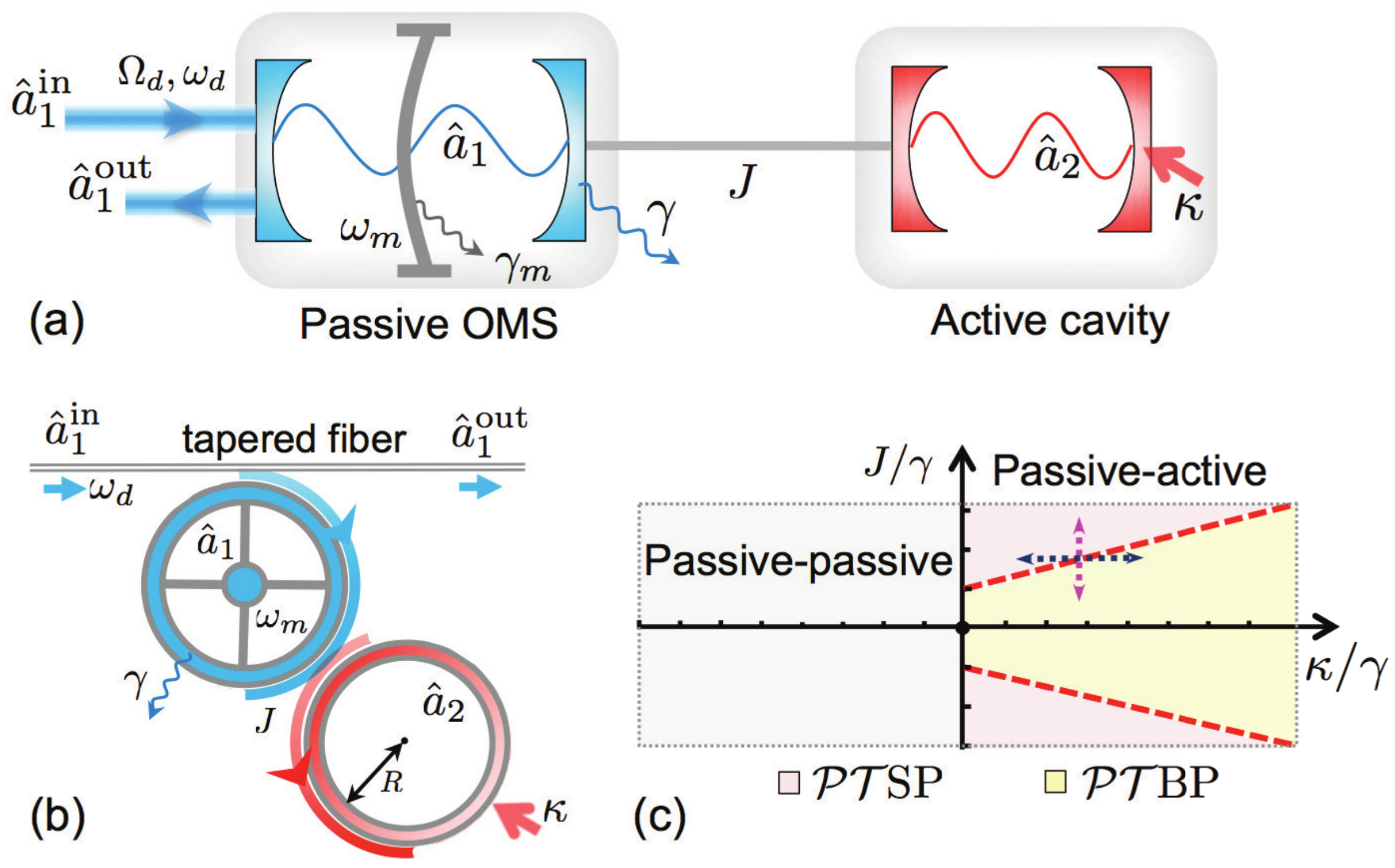}
\caption{(color online). (a) A schematic illustration of a $\mathcal{PT}$-symmetric OMS including a passive OMS (with cavity mode $\hat{a}_1$ and a mechanical mode $\hat{b}$) coupled to an active cavity $\hat{a}_2$ with tunneling strength $J$. Here $\hat{a}^{\rm in}_1$ and $\hat{a}^{\rm out}_1$ are the input and output of a driving field with frequency $\omega_d$; $\gamma$ ($\gamma_m$) and $\kappa$ are respectively the optical (mechanical) decay rate and gain rate. (b) The realization of $\mathcal{PT}$-symmetric OMS in coupled microtoridal resonators. (c) The phase-diagram in terms of $J/\gamma$ and $\kappa/\gamma$ indicates the $\mathcal{PT}$SP and $\mathcal{PT}$BP in the passive-active system, as well as the parameter regime corresponding to passive-passive system.}
\label{fig1}
\end{figure}

\emph{System.---} Consider an OMS coupled to an active cavity via optical tunneling, i.e., the $\mathcal{PT}$-symmetric OMS [see Fig.\,\ref{fig1}(a)], with the Hamiltonian
\begin{equation}
\hat{H}=\hat{H}_{c}+\hbar\omega_m\hat{b}^{\dagger}\hat{b}-\hbar g_0\hat{a}_1^{\dagger}\hat{a}_1 (\hat{b}^{\dagger}+\hat{b}),\label{H_original}
\end{equation}
where $\hat{a}_1$ ($\hat{a}_1^{\dagger}$) and $\hat{b}$ ($\hat{b}^{\dagger}$) are the annihilation (creation) operators of the passive cavity mode and mechanical mode, respectively.
The passive cavity $\hat{a}_1$ couples to an active cavity $\hat{a}_2$ (with the same resonance frequency $\omega_c$), and is driven with frequency $\omega_d$, amplitude $\Omega_d$.
Here $\Omega_d$ is related to the input power $P$ and decay rate $\gamma$ by $|\Omega_d|=\sqrt{2P\gamma/\hbar\omega_d}$. The Hamiltonian of cavity part can be written as $\hat{H}_{c}/\hbar=\Delta_{c}(\hat{a}_1^{\dagger}\hat{a}_1+\hat{a}_2^{\dagger}\hat{a}_2)-J(\hat{a}_1^{\dagger}\hat{a}_2+h.c.)-i\Omega_d(\hat{a}_1^{\dagger}-\hat{a}_1)$, with $\Delta_{c}=\omega_c-\omega_d$ in a frame rotating with $\omega_d$. The dimensionless position and momentum operators of the mechanical oscillator are defined as $\hat{x}=(1/\sqrt{2})(\hat{b}^{\dagger}+\hat{b})$ and $\hat{p}=(i/\sqrt{2})(\hat{b}^{\dagger}-\hat{b})$.
The third term in Eq.~(\ref{H_original}) describes the radiation-pressure interaction between the passive cavity and the mechanical oscillator with coupling strength $g$~\cite{CKLawPRA1995}. In principle, this $\mathcal{PT}$-symmetric OMS could be realized in many types of OMS~\cite{reviews}, such as the coupled microtoroidal resonators~\cite{Peng2014,Feng2014,Hodaei2014} [see Fig.\,\ref{fig1}(b)].

To explore the nonlinear dynamics of system, we employ the semiclassical Langevin equations of motion (setting $o=\langle\hat{o}\rangle$, $o$ is an any optical or mechanical operator)
\begin{subequations}\label{pt-ne}
\begin{align}
\dot{x}& = \omega_m p,
\\
\dot{p}& = -(\gamma_m/2) p-\omega_m x+\sqrt{2}g_0|a_1|^2,
\\
\dot{a}_1& = (-i\Delta_c-\gamma/2)a_1 + iJa_2 + i\sqrt{2}g_0a_1x + \Omega_d,\label{pt-1}
\\
\dot{a}_2& = (-i\Delta_c+\kappa/2)a_2 + iJa_1.\label{pt-2}
\end{align}
\end{subequations}
Here the quantum correlations of photon-phonon has been safely ignored in the semiclassical approximation, which is valid in the concerned weak-coupling regime, i.e., $g_0/\gamma\ll1$~\cite{Carmon2007,Bakemeier2015}. Equations~(\ref{pt-ne}) show that the intracavity field intensity and the mechanical deformation influence each other during evolution via the optomechanical interaction. Generally, this can lead to a chaotic motion of the optical and mechanical modes at a very high-power laser-driving~\cite{Carmon2007,Bakemeier2015}. However, as shown in Fig.\,\ref{fig1}(c), the present coupled OMS features a $\mathcal{PT}$-symmetry phase transition when $J/\gamma$ or $\kappa/\gamma$ passes through the EP, i.e., $J=(\gamma+\kappa)/4$, where the eigenvalues and the corresponding eigenstates of system coalesce. This property might influence the nonlinear dynamics of system significantly. Here the EP is obtained by diagonalizing the coefficient matrix of Eqs.\,(\ref{pt-1}),\,(\ref{pt-2}) under the condition of ignoring the optomechanical-interaction-induced shift of EP, which is valid in the weak-coupling regime $g_0/\gamma\ll1$~\cite{Graefe2010}.
\begin{figure}
\includegraphics[width=4.5cm]{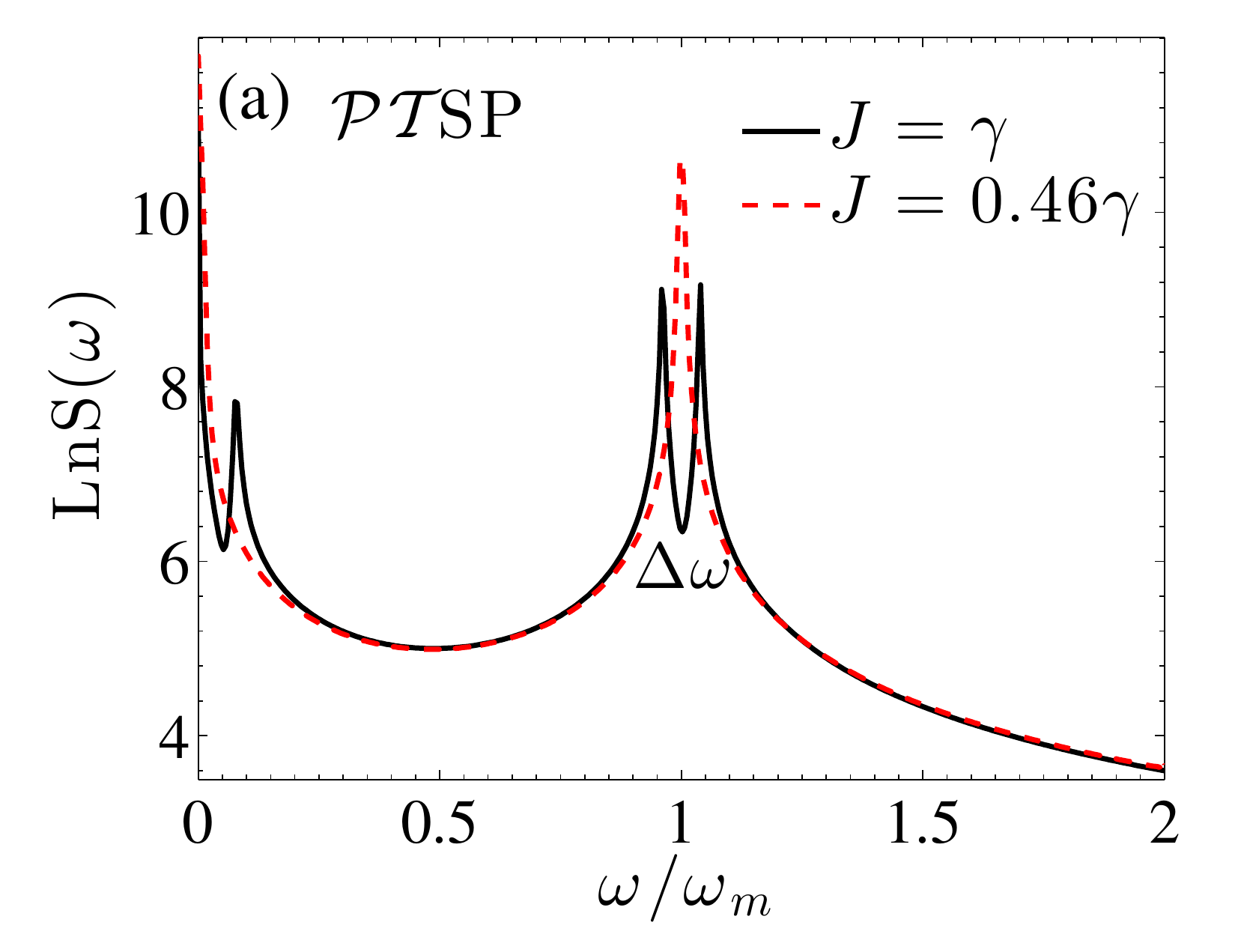}\includegraphics[width=4.5cm]{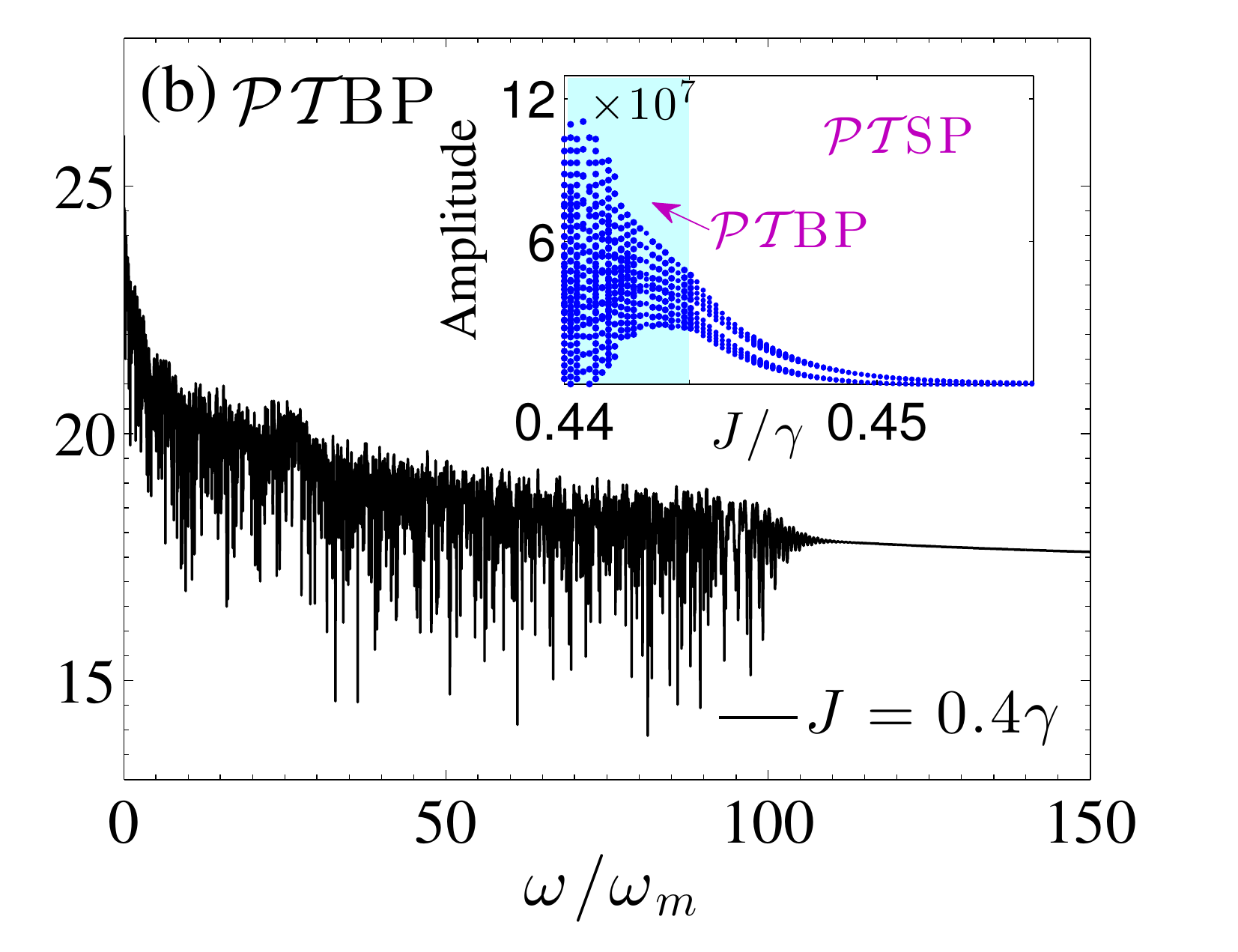}
\includegraphics[width=4.5cm]{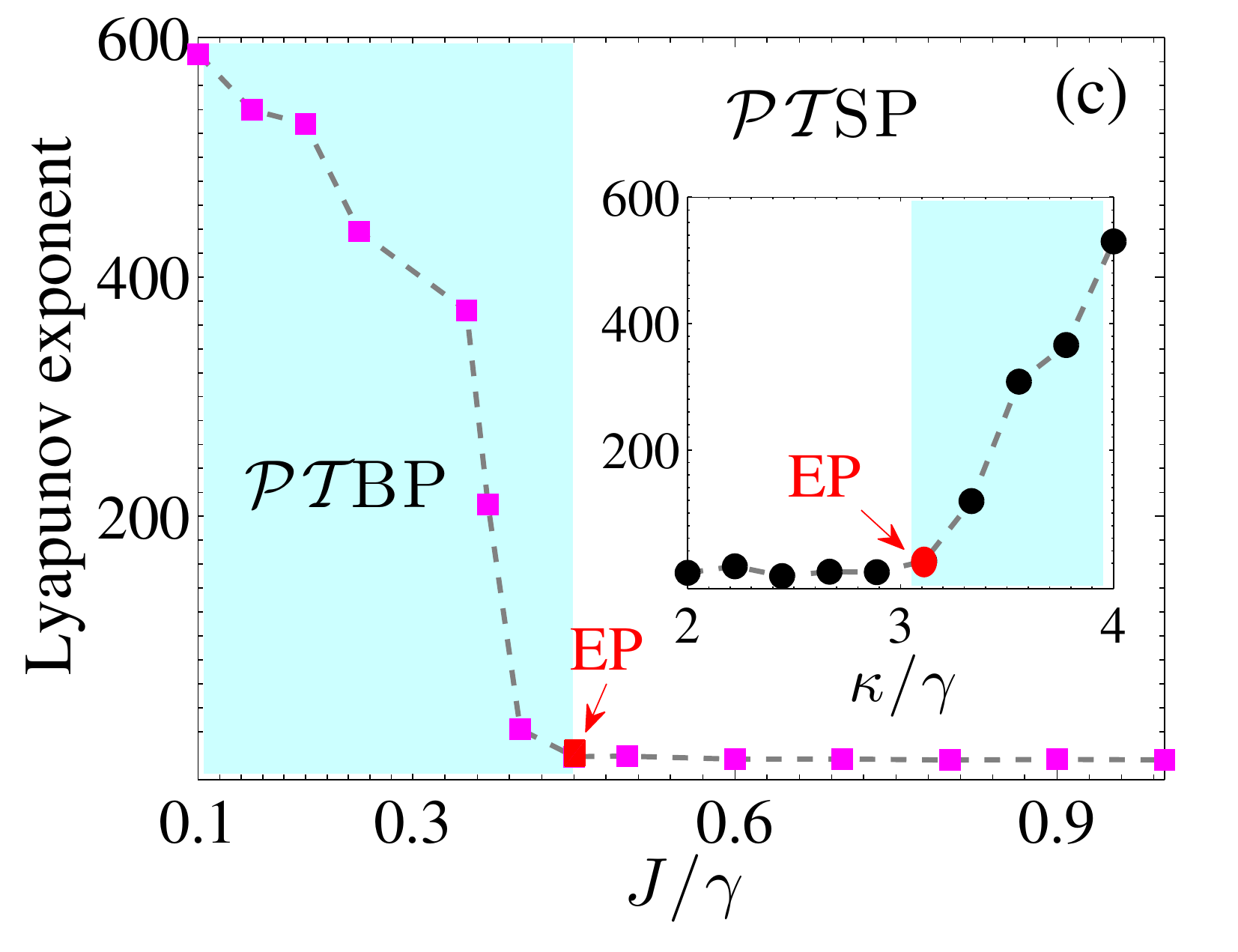}\includegraphics[width=4.5cm]{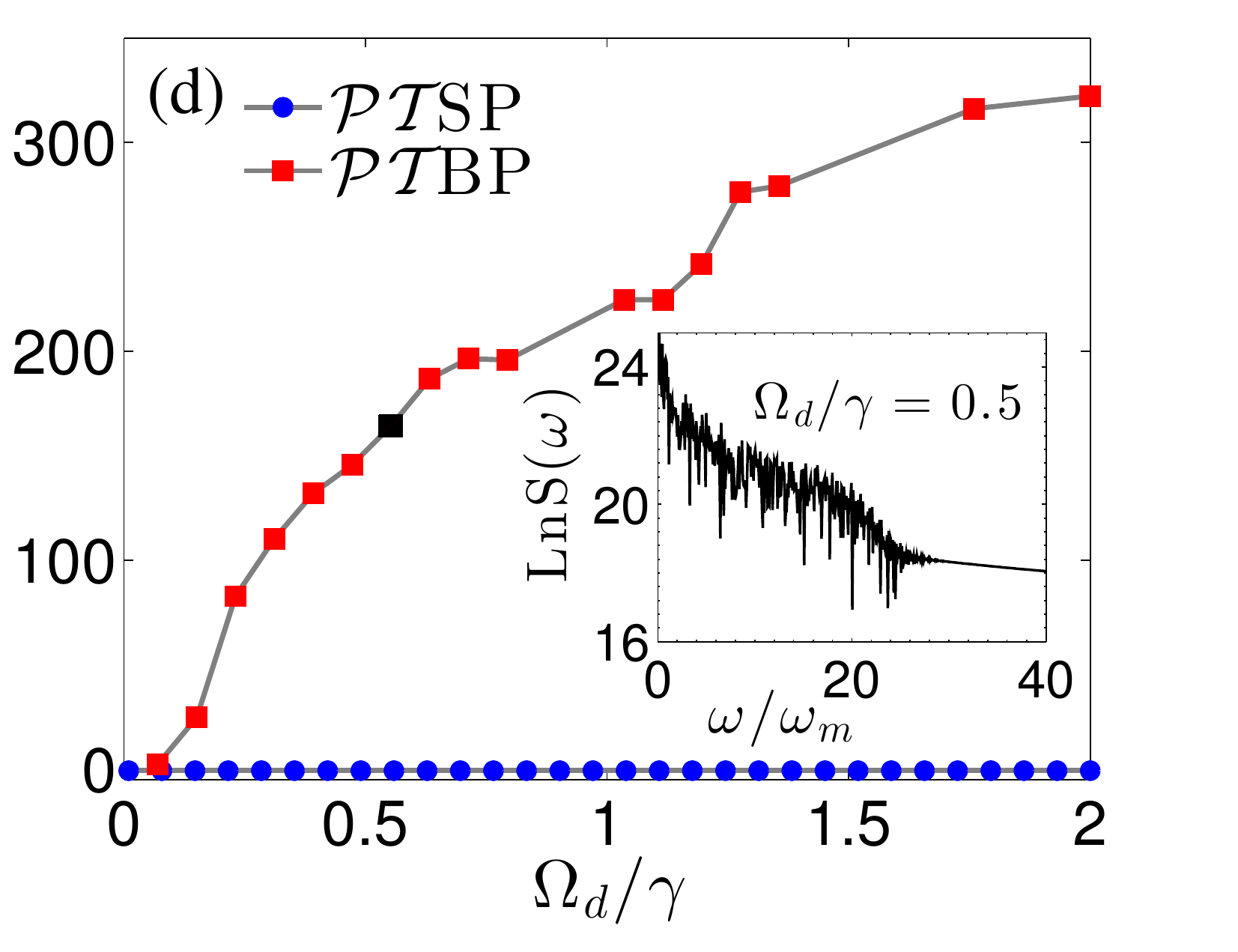}
\caption{(color online). Power spectrum ${\rm Ln}{\rm S}(\omega)$ of the intracavity field $I_1=|a_1|^2$ in $\mathcal{PT}$SP (a) and $\mathcal{PT}$BP (b). The inset of (b): bifurcation diagram versus $J/\gamma$. Lyapunov exponent versus (c) $J/\gamma$, inset of (c) $\kappa/\gamma$, and (d) $\Omega_d/\gamma$ corresponding to a fixed time interval with 0.5\,$\mu$s. 
The shadowed blue region of the inset of (b) and (c) correspond to $\mathcal{PT}$BP, and the red arrows indicate the EPs. The inset of (d) corresponds to a case of weak driving, i.e., the black square in (d). According to recent microcavity experiments~\cite{Peng2014}, the parameters are $\omega_c=190$ THz, $\omega_m/2\pi=23$ MHz, $\gamma/2\pi=1$ MHz, $g_0=7.4\times10^{-5}\gamma$, $\gamma_m/2\pi=0.038\gamma$, $\kappa=0.8\gamma$, $\Delta_c=\omega_m$, and (a-c) $P$=1\,$\mu$W (or $\Omega_d/\gamma\approx4000$), (d) $J=\gamma$ and $J=0.2\gamma$ correspond to the blue circle and red square, respectively.}
\label{fig2}
\end{figure}

Governed by Eqs.\,(\ref{pt-ne}), the evolution trajectory of system is specified by the initial condition. To characterize the stochastic properties of system, we linearize Eqs.\,(\ref{pt-ne}) and introduce the evolution of a perturbation $\vec{\delta}=(\delta\!x,\delta\!p,\delta\!a_{1r},\delta\!a_{1i},\delta\!a_{2r},\delta\!a_{2i})$, i.e., $\dot{\vec{\delta}}= \textbf{M}\vec{\delta}$ with coefficient matrix
\begin{equation}
\!\!\!\!\textbf{M}\!\!=\!\!\left(
\begin{array}{cccccc}
\!\!\!\!0\!\!&\!\!\!\omega_m\!&\!0&0\!\!&\!0\!&0\!\!\\
\!\!\!\!-\!\omega_m\!\!\!&\!\!\!-\!\gamma_m/2\!&\!2\sqrt{2}g_0a_{1r}&\!2\sqrt{2}g_0a_{1i}\!\!&\!0\!&0\!\!\\
\!\!\!\!-\!\sqrt{2}g_0a_{1i}\!\!\!&\!\!\!0\!&\!-\!\gamma/2&\!\!\Delta_c\!\!-\!\!\sqrt{2}g_0x\!\!\!&\!0\!&-\!J\!\!\\
\!\!\!\!\sqrt{2}g_0a_{1r}\!\!\!&\!\!\!0\!&\!-\!\Delta_c\!\!+\!\!\sqrt{2}g_0x&\!-\!\gamma/2\!\!&\!J\!&0\!\!\\
\!\!\!\!0\!\!\!&\!\!\!0\!&\!0&-\!\!J\!\!&\!\kappa/2\!&\Delta_c\!\!\\
\!\!\!\!0\!\!\!&\!\!\!0\!&\!J&\!0\!\!&\!-\!\Delta_c\!&\kappa/2\!\!
\end{array}
\right)\!\!,\nonumber\!\!\!\label{M}
\end{equation}
which characterizes the divergence of nearby trajectories in phase space. Here $\delta\!a_{kr}$ and $\delta\!a_{ki}$ are respectively the perturbation of the real part and imaginary part of $a_k$ ($k=1,2$).

\begin{figure}
\includegraphics[width=8.6cm]{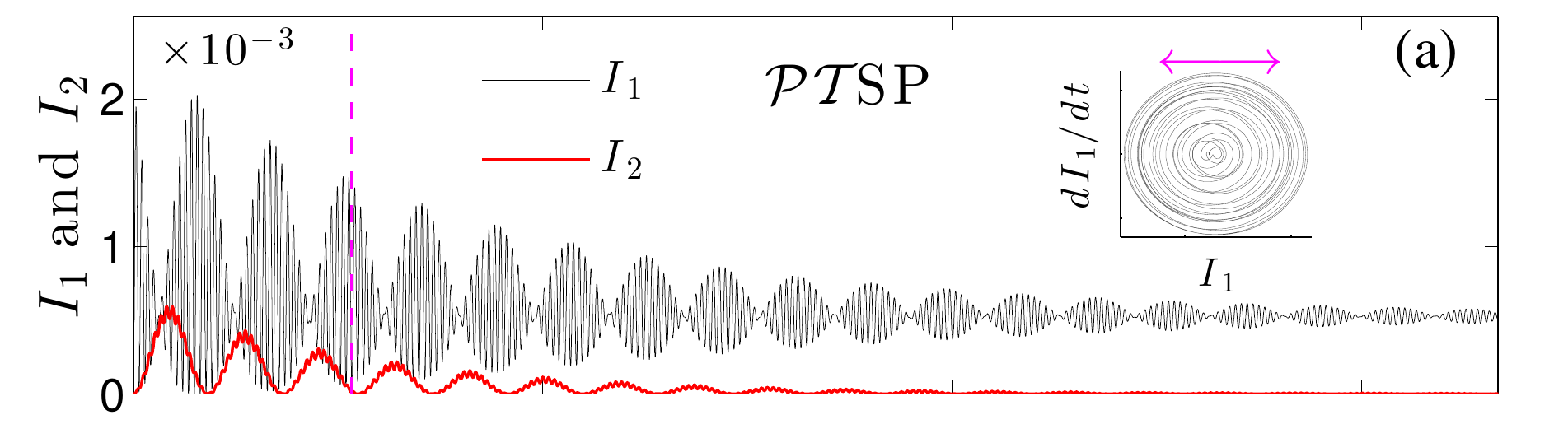}
\includegraphics[width=8.6cm]{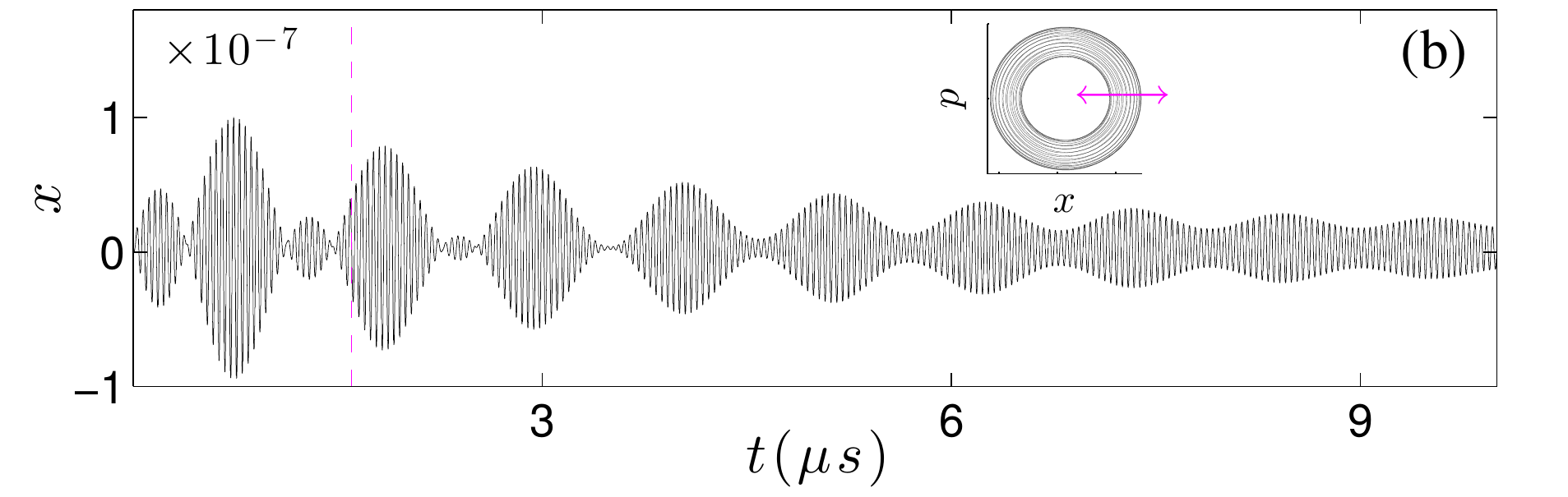}
\caption{(color online). Evolutions of (a) the intracavity intensities $I_1$ (passive cavity), $I_2$ (active cavity), and (b) the dimensionless mechanical displacement $x$ in $\mathcal{PT}$SP, i.e., $J=\gamma$.
The insets: the trajectories of (a) the optical mode, i.e., the first derivation of $I_1$ versus $I_1$ and (b) the mechanical mode, i.e., $p$ versus $x$, in phase space corresponding to a fixed time interval $8\,\mu$s\,$\rightarrow9\,\mu$s. The purple dashed lines indicates the amplitudes of $I_1$, $I_2$ and $x$ at 1.6 $\mu$s. The purple arrows indicate the approximate evolution direction of the trajectories. The parameters are same as Fig.\,\ref{fig2} except for $P\approx0.02$ pW, i.e., $\Omega_d=0.5\gamma$.}
\label{fig3}
\end{figure}
\emph{$\mathcal{PT}$-symmetry-breaking-induced chaos.---}
By numerically solving Eqs.\,(\ref{pt-ne}), in Figs.\,\ref{fig2}(a) and \ref{fig2}(b) we respectively present the power spectrum of the field intensity $I_1=|a_1|^2$ in $\mathcal{PT}$SP and $\mathcal{PT}$BP, corresponding to a fixed driving power $P=1\,\mu$W and time interval $8\rightarrow9\,\mu$s.
On one hand, it shows that the cavity mode is periodically modulated by the mechanical oscillation with frequency $\omega_m$ in $\mathcal{PT}$SP at a weak tunnelling coupling $J=0.46\gamma$ [very close to the EP $J=(\gamma+\kappa)/4=0.45\gamma$]. When $J$ is increased so strong that it could be comparable with (or larger than) the optical linewidth $\gamma$, a resolved normal-model-splitting with width $\Delta\omega=\sqrt{16J^2-(\gamma+\kappa)^2}/2$ appears around the center frequency and mechanical sideband due to the strong tunneling effect between the active and passive cavities. On the other hand, decreasing $J/\gamma$ pushes the system into the $\mathcal{PT}$BP, and the power spectrum becomes a continuum characterizing the emergence of chaotic motion. Here the driving power 1\,$\mu$W is well below the threshold of normal optomechanical chaos for more than 3 order due to the field-localization-enhanced nonlinearity. Moreover, in the inset of Fig.\,\ref{fig2}(b), we present the bifurcation diagram (corresponding to the time interval $8\rightarrow9\,\mu$s) as the control parameter $J/\gamma$ is varied, which shows that the system dynamics experiences regular, period-doubling bifurcation to chaos behaviors from $\mathcal{PT}$SP to $\mathcal{PT}$BP.

In Figs.\,\ref{fig2}(c) and \ref{fig2}(d), we present the dependence of Lyapunov exponent~\cite{Oseledec1968} on $J/\gamma$, $\kappa/\gamma$ and $\Omega_d/\gamma$. The Lyapunov exponent is defined by the logarithmic slope of the perturbation $\delta I_1$ ($\delta I_1=|a_1+\delta a_1|^2-|a_1|^2$) versus time $t$, and characterizes the separation of trajectories for identical systems with infinitesimally close initial condition. It is obtained by numerically solving Eq.\,(\ref{pt-ne}) and the perturbation equation $\dot{\vec{\delta}}= \textbf{M}\vec{\delta}$~\cite{Carmon2007}. It clearly shows that the chaotic motion appears when one tunes the system parameters so that $\mathcal{PT}$ symmetry is broken, which provides an alternative method to control chaotic dynamics with $\mathcal{PT}$-symmetry phase transition.  Moreover, Fig.\,\ref{fig2}(d) also shows that the chaos allows an ultralow threshold, such as the {\it weak-driving} regime $\Omega_d=0.5\gamma$ corresponding to $P=0.02\,$pW for microcavity system~\cite{Peng2014}. Note that, the oscillation amplitude of cavity intensity  is not very large in the $\mathcal{PT}$SP. However the semiclassical equation (\ref{pt-ne}) is still valid, which is ensured by the bad-cavity regime, i.e., $g_0/\kappa\ll1$ and by using classical driving~\cite{Carmon2007,Bakemeier2015}. Hence the Lyapunov exponent in Fig.\,\ref{fig2} obtained from Eq.\,(\ref{pt-ne}) can be trusted, which is also consistent with the evolution of system Fig.\,\ref{fig3}.  
 
\begin{figure}
\includegraphics[width=8.2cm]{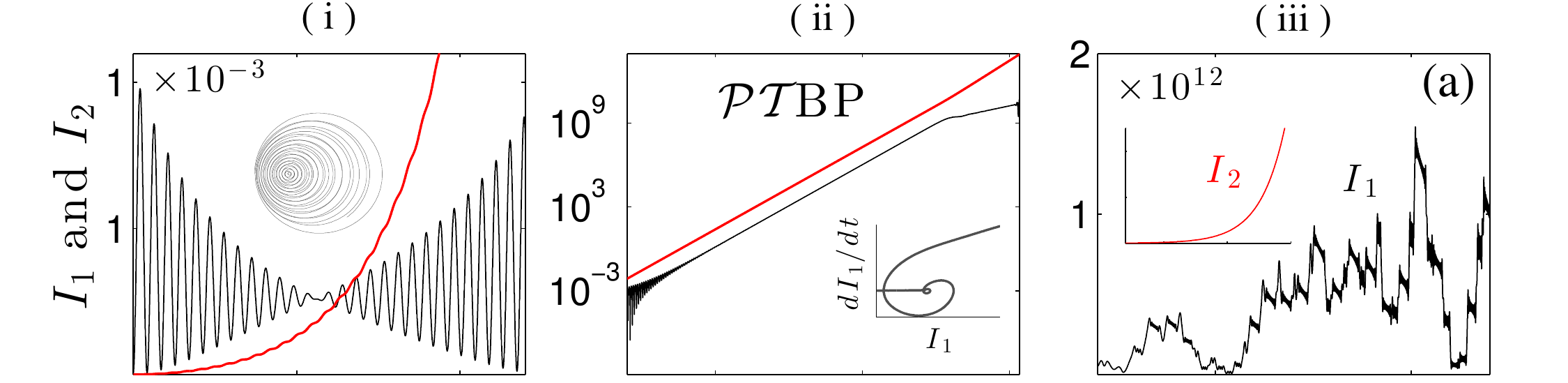}
\includegraphics[width=8.2cm]{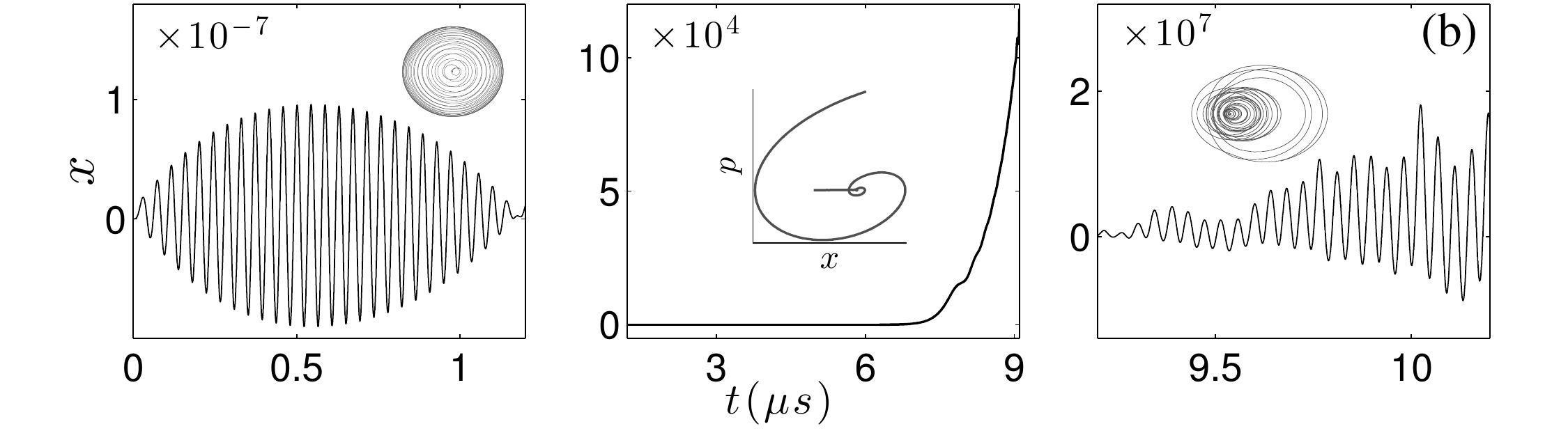}
\includegraphics[width=8.2cm]{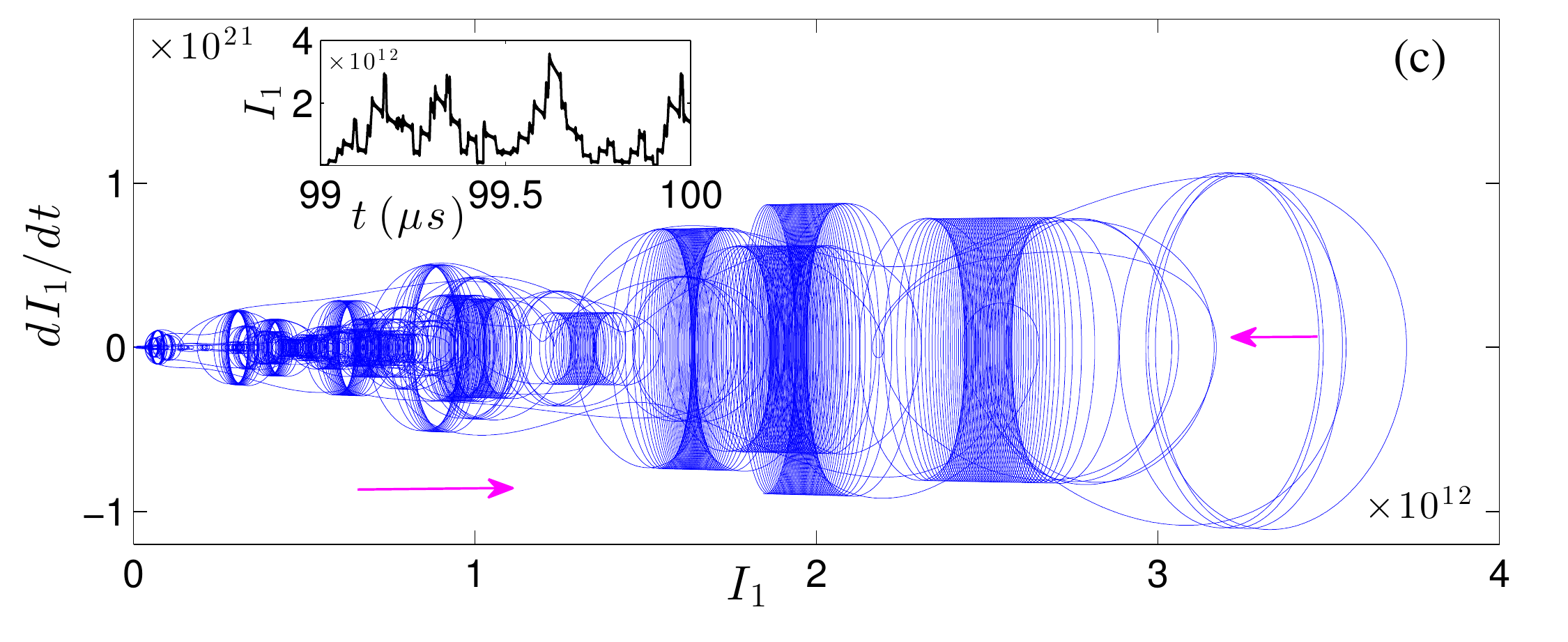}
\caption{(color online). Evolutions of (a) $I_1$, $I_2$, and (b) $x$ in $\mathcal{PT}$BP, i.e., $J=0.2\gamma$.
The optical and mechanical trajectories in phase spaces are presented in the insets of (a,b). 
(c) The optical trajectory corresponding to time interval (iii). The inset of (c) presents the chaos dynamics in the long-time limit. 
The purple arrow in (c) indicates the approximate evolution direction of trajectory. The parameters are same as Fig.\,\ref{fig3}.}
\label{fig4}
\end{figure}

\emph{Dynamical trajectories in different $\mathcal{PT}$ phases.---}
To understand the ultralow threshold chaos in $\mathcal{PT}${\rm BP}, in Figs.\,\ref{fig3} and \ref{fig4}, we present the evolutions of the intracavity intensities $I_1$, $I_2$ and the mechanical displacement $x$ in $\mathcal{PT}${\rm SP} and $\mathcal{PT}${\rm BP}, respectively.

Figures~\ref{fig3} show the amplitude oscillations in $I_1$, $I_2$ and $x$, characterizing the strong energy exchanges between the passive cavity and the active cavity as well as the mechanical oscillator (see the purple-dashed line). This ultimately leads to the periodic dynamics of the optical and mechanical modes at a low driving-power in $\mathcal{PT}$SP. Accordingly, the optical and mechanical trajectories are limited into the regular circles with periodically varying radius with evolution (see the insets of Fig.\,\ref{fig3}). Physically, this is because, in $\mathcal{PT}$SP, the cavity-tunnelling effects characteristized by $J$ is stronger than the intracavity localization effects characteristized by $(\kappa+\gamma)/4$. Without strong optical driving, the weak intracavity intensity $I_1$ cannot induce enough optomechanical nonlinearity to lead to a chaotic motion.

Figures~\ref{fig4} show that, in $\mathcal{PT}$BP, the optical fields are localized respectively into their cavities (no clear energy oscillation between two cavities), and the chaotic motions of the optical and mechanical modes appear after a proper evolution even in the {\it weak-driving} regime, i.e., $\Omega_d/\gamma=0.5$. Comparing with the case of $\mathcal{PT}$SP, here the evolution of $I_1$ is more complicated, which can be divided into three distinguishable processes. (i) An analogous collapse and revival appears due to the interplay between the optical loss and driving. The trajectories in phase space feature a set of limited circles accordingly.  (ii) It presents an exponential growth almost without oscillation with evolution, which demonstrates an enormous energy accumulation during this time interval. The approximate logarithmic-spiral-trajectory in phase space also indicate this property. Physically, it comes from the unidirectional energy transfer from the active cavity to the passive cavity in $\mathcal{PT}$BP, which also can submerge a probe laser injected from the active cavity and lead to a non-reciprocity transmission~\cite{Peng2014}.  (iii) Chaotic motions of the optical and mechanical oscillations emerges and lasts with evolution. The trajectories in phase spaces become very complicated [looks like a three-dimensional spirals, see Fig.\,\ref{fig4}(c)]. Accordingly, the mechanical trajectory goes trough similar processes [see Fig.\,\ref{fig4}(b)]. Note that the present chaos is bounded in the chaos regime due to the enhanced optomechanical nonlinearity, as shown in the inset of Fig.\,\ref{fig4}(c), i.e., the evolution of system in the long-time limit. Based on the above, one can conclude that the essence of $\mathcal{PT}$-symmetry-breaking chaos is the localization-induced dynamical-intensity-accumulation in $\mathcal{PT}$BP. 
Therefore this chaos can be triggered even if an arbitrary small amount of power is applied to the system. Because strong intercavity intensity can be accumulated during the processes (i) and (ii) due to the effect of field localization. It can induce enough large optomechanical-nonlinearity required by chaotic motion.

\begin{figure}
\includegraphics[width=8.6cm]{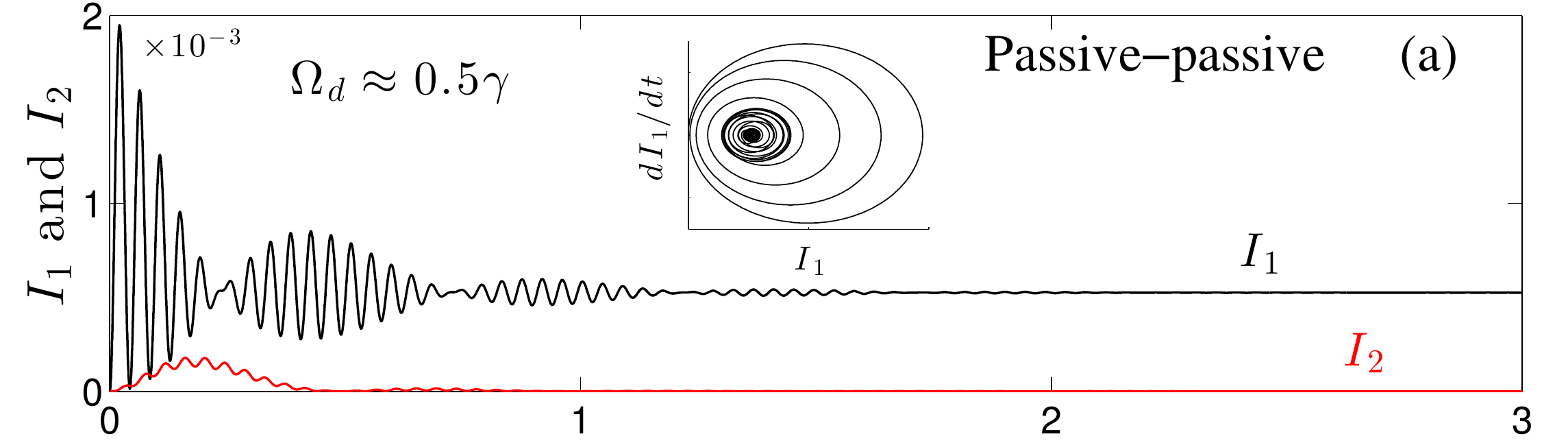}
\includegraphics[width=8.6cm]{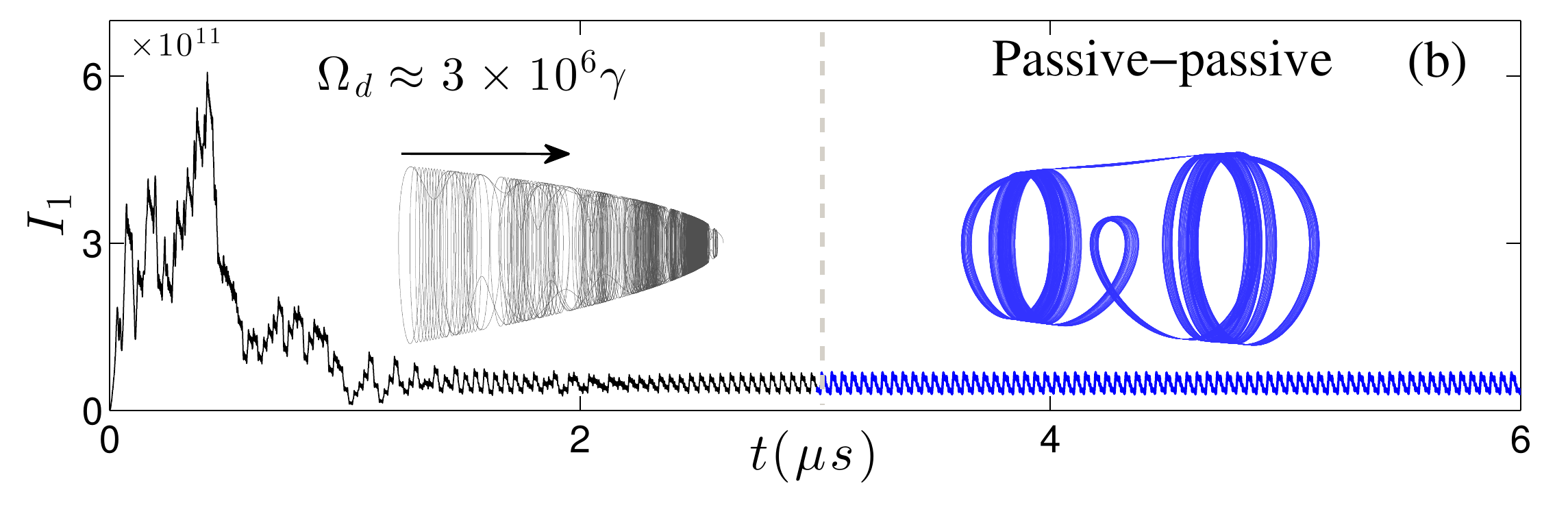}
\includegraphics[width=8.6cm]{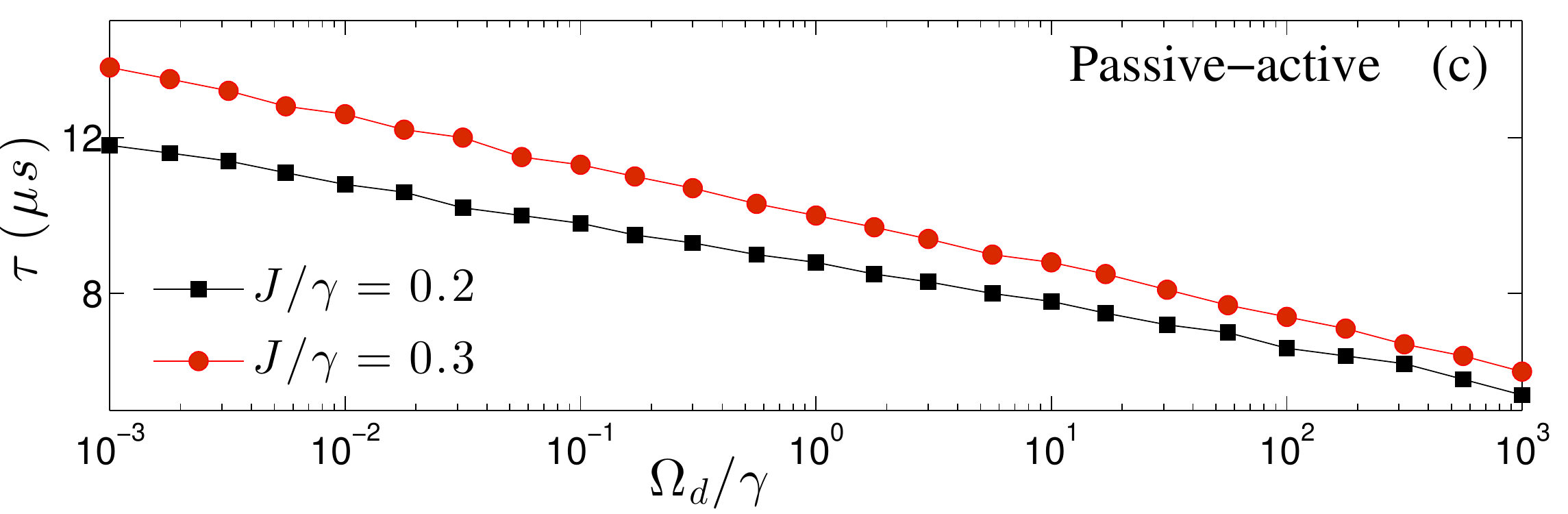}
\caption{(color online). The evolutions of $I_1$ and $I_2$ in the passive-passive system when (a) $P\approx0.02$ pW or $\Omega_d\approx0.5\gamma$ and (b) $P\approx500$ mW or $\Omega_d\approx3\times10^6\gamma$. The insets present the corresponding optical trajectories in phase space. The black arrow indicates the approximate evolution direction of trajectory. (c) The approximate starting time $\tau$ of chaotic motion versus driving strength $\Omega_d/\gamma$ in the active-passive system.
The parameters are same as Fig.\,\ref{fig2} except for (a,b) $\kappa=-0.8\gamma$.}
\label{fig5}
\end{figure}

\emph{Comparison with the normal chaotic dynamics.---}
The above $\mathcal{PT}$-symmetric dynamics is quite different from the nonlinear dynamics in the normal passive-system, especially in the chaotic regime. When $\kappa<0$ the $\mathcal{PT}$-symmetric system returns to a passive-passive system [see Fig.\,\ref{fig1}(c)], whose dynamical evolutions are presented in Figs.\,\ref{fig5}(a) and (b). The comparison between Fig.\,\ref{fig3}(a) and Fig.\,\ref{fig5}(a) shows an enhanced optical tunneling phenomenon in $\mathcal{PT}$-symmetric system. Moreover, Fig.\,\ref{fig5}(b) shows that very strong optical driving (the order of mW for microcavity OMS) is required to obtain chaotic motion in passive-passive system. The chaotic motion appears at the beginning time and disappears with evolution due to the optical loss. Accordingly there is a transition from chaotic trajectory to regular trajectory in phase space, as shown in the insets of Fig.\,\ref{fig5}(b). Then, comparing Fig.\,\ref{fig4}(a) with Fig.\,\ref{fig5}(b), one can see that the $\mathcal{PT}$-symmetry-breaking chaos is quite different from the normal optomechanical chaos. (1) It emerges after going through two distinguishable processes, and lasts with evolution. (2) It allows an ultralow threshold (even is thresholdless), i.e., in principle, it can be triggered once a non-zero-power driving laser is employed and lasts for a period of time until enough nonlinearity is reached. The driving strength only resolves the starting time of chaotic motion, as shown in Fig.\,\ref{fig5}(c). Strong optical driving makes the chaos arise more quickly. 

\emph{Conclusions.---} We have investigated the nonlinear dynamics of an OMS coupled to an active cavity, which features a controllable $\mathcal{PT}$-symmetry phase transition. We showed that, by tuning the tunnelling-to-loss or gain-to-loss rate, we can obtain an optomechanical chaos induced by the $\mathcal{PT}$-symmetry breaking. Moreover we also presented the dynamical trajectories of system in $\mathcal{PT}$SP and $\mathcal{PT}$BP, respectively. The increasing optomechanical nonlinearity with evolution in $\mathcal{PT}$BP makes the chaos features an ultralow threshold, which is required for many secret communication schemes. This study provides a promising route for controlling nonlinear dynamics, especially the generation of chaos, with the concept of $\mathcal{PT}$ symmetry. It opens up new avenues for the study of ultralow-power-laser triggered chaos. Conversely, it also offers a method to distinguish the $\mathcal{PT}$SP and $\mathcal{PT}$BP via the experimentally detectable optical-spectrum.

\emph{Acknowledgements.---} XYL, HJ thanks Dr. J. Zhang, Dr. S.K. \"{O}zdemir, Prof.~L. Yang and Prof.~F. Nori for valuable discussions. XYL is supported by NSFC-11374116. HJ is supported by the NSFC (Contracts No. 11274098 and 11474087). YW is supported by NSFC-11375067, and the National Basic Research Program of China (Contracts No. 2012CB922103).


\begin{thebibliography}{99}
\bibitem{reviews} For recent reviews, see, e.g., M. Aspelmeyer, T.J. Kippenberg, F. Marquardt, Rev. Mod. Phys. \textbf{86}, 1391 (2014);
M. Aspelmeyer, P. Meystre and K. Schwab, Physics Today \textbf{65}(7), 29 (2012); P. Meystre, Ann. Phys. (Berlin, Ger.) \textbf{525}, 215 (2013);
F. Marquardt and S.M. Girvin, Physics \textbf{2}, 40 (2009); T.J. Kippenberg and K.J. Vahala, Science \textbf{321}, 1172 (2008).

\bibitem{Bose1997} S. Bose, K. Jacobs, and P.L. Knight, Phys. Rev. A \textbf{56}, 4175 (1997);
W. Marshall, C. Simon, R. Penrose, and D. Bouwmeester, Phys. Rev. Lett. \textbf{91}, 130401 (2003).

\bibitem{Rabl2011} P. Rabl, Phys. Rev. Lett. \textbf{107}, 063601 (2011);
A. Nunnenkamp, K. B{\o}rkje, and S.M. Girvin, Phys. Rev. Lett. \textbf{107}, 063602 (2011).

\bibitem{Liao2012} J.Q. Liao, H. K. Cheung, and C.K. Law, Phys. Rev. A \textbf{85}, 025803 (2012); X.-W. Xu, H. Wang, J. Zhang, and Y. X. Liu, Phys. Rev.
A \textbf{88}, 063819 (2013).

\bibitem{Ludwig2012} M. Ludwig, A. H. Safavi-Naeini, O. Painter, and F. Marquardt,
Phys. Rev. Lett. \textbf{109}, 063601 (2012); K. Stannigel, P. Komar, S.J.M. Habraken, S.D. Bennett,
M.D. Lukin, P. Zoller, and P. Rabl, Phys. Rev. Lett. \textbf{109}, 013603 (2012).

\bibitem{Nori2013} X.-Y. L\"{u}, W.-M. Zhang, S. Ashhab, Y. Wu, and F. Nori,
Sci. Rep. \textbf{3}, 2943 (2013); M.-A. Lemonde, N. Didier, and A.A. Clerk, Phys. Rev.
Lett. \textbf{111}, 053602 (2013).

\bibitem{Xuereb2012} A. Xuereb, C. Genes, and A. Dantan, Phys. Rev. Lett.
\textbf{109}, 223601 (2012); C. Genes, A. Xuereb, G. Pupillo, and A. Dantan, Phys. Rev. A \textbf{88}, 033855 (2013).

\bibitem{Nori2015} X.-Y. L\"{u}, J.-Q. Liao, L. Tian, F. Nori, Phys. Rev. A \textbf{91}, 013834 (2015).

\bibitem{Kippenberg2005} T.J. Kippenberg, H. Rokhsari, T. Carmon, A. Scherer, and K.J. Vahala, Phys. Rev. Lett. \textbf{95}, 033901 (2005); T. Carmon, H. Rokhsari, L. Yang, T.J. Kippenberg, and
K.J. Vahala, Phys. Rev. Lett. \textbf{94}, 223902 (2005).

\bibitem{Marquardt2006} F. Marquardt, J.G.E. Harris, and S.M. Girvin, Phys. Rev. Lett. \textbf{96}, 103901 (2006).

\bibitem{Metzger2008} C. Metzger et al., Phys. Rev. Lett. \textbf{101}, 133903 (2008).

\bibitem{Zaitsev2011} S. Zaitsev, A. K. Pandey, O. Shtempluck, and E. Buks, Phys. Rev. E \textbf{84}, 046605 (2011).

\bibitem{Carmon2007} T. Carmon, M.C. Cross, and K.J. Vahala, Phys. Rev. Lett. \textbf{98}, 167203 (2007).

\bibitem{Meystre2010} K. Zhang, W. Chen, M. Bhattacharya, and P. Meystre, Phys. Rev. A \textbf{81}, 013802 (2010).

\bibitem{Ma2014} J. Ma et al., Phys. Rev. A \textbf{90}, 043839 (2014).

\bibitem{Bakemeier2015} L. Bakemeier, A. Alvermann, and H. Fehske, Phys. Rev. Lett. \textbf{114}, 013601 (2015).

\bibitem{Xiong2014} H. Xiong, L.-G. Si,  X.-Y. L\"{u}, X. Yang, Y. Wu, Annals of Physics \textbf{349}, 43 (2014).

\bibitem{Gleeson} J.T. Gleeson, Appl. Phys. Lett. \textbf{81}, 1949 (2002).

\bibitem{Sivaprakasam} S. Sivaprakasam and K.A. Shore, Opt. Lett. \textbf{24}, 466 (1999).

\bibitem{VanWiggeren} G.D. VanWiggeren and R. Roy, Science \textbf{279}, 1198 (1998).

\bibitem{Sciamanna} M. Sciamanna and K.A. Shore, Nature Photonics \textbf{9}, 151 (2015). 

\bibitem{Miller} D. Miller, Proc. IEEE \textbf{97}, 1166 (2009).

\bibitem{Ellis} B. Ellis et al., Nature Photonics \textbf{5}, 297 (2011).

\bibitem{Bender1998} C.M. Bender and S. Boettcher, Phys. Rev. Lett. \textbf{80}, 5243 (1998); C.M. Bender, Rep. Prog. Phys. \textbf{70}, 947 (2007);

\bibitem{Makris2008} K.G. Makris, R. El-Ganainy, D.N. Christodoulides, Z.H. Musslimani, Phys. Rev. Lett. \textbf{100}, 103904 (2008).

\bibitem{Klaiman2008} S. Klaiman, U. G\"{u}nther, N. Moiseyev, Phys. Rev. Lett. \textbf{101}, 080402 (2008).

\bibitem{West2010} C.T. West, T. Kottos, and T. Prosen, Phys. Rev. Lett. \textbf{104}, 054102 (2010).

\bibitem{Guo2009} A. Guo et al., Phys. Rev. Lett. \textbf{103}, 093902 (2009).

\bibitem{Makris2010} C.E. R\"{u}ter et al., Nat. Phys. \textbf{6}, 192 (2010).

\bibitem{Chong2011} Y.D. Chong, L. Ge, A.D. Stone, Phys. Rev. Lett. \textbf{106}, 093902 (2011).

\bibitem{Miroshnichenko2011} A.E. Miroshnichenko, B.A. Malomed, Y.S. Kivshar, Phys. Rev. A \textbf{84}, 012123 (2011).

\bibitem{Lin2011} Z. Lin et al., Phys. Rev. Lett. \textbf{106}, 213901 (2011).

\bibitem{Regensburger2012} A. Regensburger, C. Bersch, M.-A. Miri, G. Onishchukov, D.N. Christodoulides, and U. Peschel, Nature \textbf{488}, 167 (2012).

\bibitem{Feng2013} L. Feng et al., Nat. Mater. \textbf{12}, 747 (2013).

\bibitem{Peng2014} B. Peng et al., Nat. Phys. \textbf{10}, 394 (2014); B. Peng et al., Science \textbf{346}, 328 (2014); 
L. Chang et al., Nature Photonics \textbf{8}, 524 (2014).

\bibitem{Feng2014} L. Feng, Z. J. Wong, R.-M. Ma, Y. Wang, X. Zhang, Science \textbf{346}, 972 (2014).

\bibitem{Hodaei2014} H. Hodaei, M.-A. Miri, M. Heinrich, D.N. Christodoulides, M. Khajavikhan, Science \textbf{346}, 975 (2014).

\bibitem{Jing2014} H. Jing et al., Phys. Rev. Lett. \textbf{113}, 053604 (2014); H. Jing et al., Sci. Rep. 5, 9663 (2015).

\bibitem{Xie2014} X. Li and X.-T. Xie, Phys. Rev. A \textbf{90}, 033804 (2014).

\bibitem{footnote} This chaos belongs to the classical chaos, which is quite different from the dynamical wave localization (or quantum chaos) in the celebrated kicked rotor~\cite{West2010}.

\bibitem{CKLawPRA1995} C.K. Law, Phys. Rev. A \textbf{51}, 2537 (1995).

\bibitem{Graefe2010} E.-M. Graefe, H.J. Korsch, and A.E. Niederle, Phys. Rev. A \textbf{82}, 013629 (2010);
H. Ramezani, T. Kottos, R. El-Ganainy, and D. N. Christodoulides, Phys. Rev. A \textbf{82}, 043803 (2010).

%\bibitem{Feigenbaum} M.J. Feigenbaum, J. Stat. Phys. \textbf{21}, 669 (1979).

\bibitem{Oseledec1968} V.I. Oseledec, Trans Mosc. Mat. Soc. \textbf{19}, 179 (1968).

\end{thebibliography}
\end{document}